\documentclass[aps,prl,reprint,superscriptaddress]{revtex4-2}

\usepackage{graphicx}
\usepackage{dcolumn}
\usepackage{bm}
\usepackage{textgreek}

\begin{document}

\title{\textbf{Collimated Hard X-Rays from Hybrid Laser and Plasma Wakefield Accelerators} 
}

\author{Hong Zhang}
\affiliation{State Key Laboratory of Ultra-intense Laser Science and Technology, Shanghai Institute of Optics and Fine Mechanics, Chinese Academy of Sciences, Shanghai 201800, China}
\affiliation{School of Optoelectronics, University of Chinese Academy of Sciences, Beijing 100049, China }

\author{Jianmeng Wei}
\affiliation{State Key Laboratory of Ultra-intense Laser Science and Technology, Shanghai Institute of Optics and Fine Mechanics, Chinese Academy of Sciences, Shanghai 201800, China}

\author{Mengyuan Chu}
\affiliation{State Key Laboratory of Ultra-intense Laser Science and Technology, Shanghai Institute of Optics and Fine Mechanics, Chinese Academy of Sciences, Shanghai 201800, China}

\author{Jiale Zheng}
\affiliation{State Key Laboratory of Ultra-intense Laser Science and Technology, Shanghai Institute of Optics and Fine Mechanics, Chinese Academy of Sciences, Shanghai 201800, China}
\affiliation{School of Optoelectronics, University of Chinese Academy of Sciences, Beijing 100049, China }

\author{Zhiheng Lou}
\affiliation{State Key Laboratory of Ultra-intense Laser Science and Technology, Shanghai Institute of Optics and Fine Mechanics, Chinese Academy of Sciences, Shanghai 201800, China}

\author{Ruoxuan Ma}
\affiliation{State Key Laboratory of Ultra-intense Laser Science and Technology, Shanghai Institute of Optics and Fine Mechanics, Chinese Academy of Sciences, Shanghai 201800, China}

\author{Xizhuan Chen}
\affiliation{State Key Laboratory of Ultra-intense Laser Science and Technology, Shanghai Institute of Optics and Fine Mechanics, Chinese Academy of Sciences, Shanghai 201800, China}
\affiliation{School of Optoelectronics, University of Chinese Academy of Sciences, Beijing 100049, China }

\author{Hao Wang}
\affiliation{State Key Laboratory of Ultra-intense Laser Science and Technology, Shanghai Institute of Optics and Fine Mechanics, Chinese Academy of Sciences, Shanghai 201800, China}

\author{Gaojie Zeng}
\affiliation{State Key Laboratory of Ultra-intense Laser Science and Technology, Shanghai Institute of Optics and Fine Mechanics, Chinese Academy of Sciences, Shanghai 201800, China}

\author{Hang Guo}
\affiliation{State Key Laboratory of Ultra-intense Laser Science and Technology, Shanghai Institute of Optics and Fine Mechanics, Chinese Academy of Sciences, Shanghai 201800, China}
\affiliation{School of Optoelectronics, University of Chinese Academy of Sciences, Beijing 100049, China }

\author{Yinlong Zheng}
\affiliation{State Key Laboratory of Ultra-intense Laser Science and Technology, Shanghai Institute of Optics and Fine Mechanics, Chinese Academy of Sciences, Shanghai 201800, China}

\author{Hai Jiang}
\affiliation{State Key Laboratory of Ultra-intense Laser Science and Technology, Shanghai Institute of Optics and Fine Mechanics, Chinese Academy of Sciences, Shanghai 201800, China}

\author{Yanjie Ge}
\affiliation{State Key Laboratory of Ultra-intense Laser Science and Technology, Shanghai Institute of Optics and Fine Mechanics, Chinese Academy of Sciences, Shanghai 201800, China}
\affiliation{School of Optoelectronics, University of Chinese Academy of Sciences, Beijing 100049, China }

\author{Kangnan Jiang}
\affiliation{State Key Laboratory of Ultra-intense Laser Science and Technology, Shanghai Institute of Optics and Fine Mechanics, Chinese Academy of Sciences, Shanghai 201800, China}

\author{Runshu Hu}
\affiliation{State Key Laboratory of Ultra-intense Laser Science and Technology, Shanghai Institute of Optics and Fine Mechanics, Chinese Academy of Sciences, Shanghai 201800, China}
\affiliation{School of Optoelectronics, University of Chinese Academy of Sciences, Beijing 100049, China }

\author{Jiayi Qian}
\affiliation{State Key Laboratory of Ultra-intense Laser Science and Technology, Shanghai Institute of Optics and Fine Mechanics, Chinese Academy of Sciences, Shanghai 201800, China}

\author{Jiacheng Zhu}
\affiliation{State Key Laboratory of Ultra-intense Laser Science and Technology, Shanghai Institute of Optics and Fine Mechanics, Chinese Academy of Sciences, Shanghai 201800, China}

\author{Zongxin Zhang}
\affiliation{State Key Laboratory of Ultra-intense Laser Science and Technology, Shanghai Institute of Optics and Fine Mechanics, Chinese Academy of Sciences, Shanghai 201800, China}

\author{Yi Xu}
\affiliation{State Key Laboratory of Ultra-intense Laser Science and Technology, Shanghai Institute of Optics and Fine Mechanics, Chinese Academy of Sciences, Shanghai 201800, China}

\author{Yuxin Leng}
\affiliation{State Key Laboratory of Ultra-intense Laser Science and Technology, Shanghai Institute of Optics and Fine Mechanics, Chinese Academy of Sciences, Shanghai 201800, China}
\affiliation{School of Optoelectronics, University of Chinese Academy of Sciences, Beijing 100049, China }

\author{Song Li}
\email{Contact author: lisong@siom.ac.cn}
\affiliation{State Key Laboratory of Ultra-intense Laser Science and Technology, Shanghai Institute of Optics and Fine Mechanics, Chinese Academy of Sciences, Shanghai 201800, China}

\author{Ke Feng}
 \email{Contact author: fengke@siom.ac.cn}
\affiliation{State Key Laboratory of Ultra-intense Laser Science and Technology, Shanghai Institute of Optics and Fine Mechanics, Chinese Academy of Sciences, Shanghai 201800, China}
 
\author{Wentao Wang}
 \email{Contact author: wwt1980@siom.ac.cn}
\affiliation{State Key Laboratory of Ultra-intense Laser Science and Technology, Shanghai Institute of Optics and Fine Mechanics, Chinese Academy of Sciences, Shanghai 201800, China}
\affiliation{School of Optoelectronics, University of Chinese Academy of Sciences, Beijing 100049, China }
 
\author{Ruxin Li}
\affiliation{State Key Laboratory of Ultra-intense Laser Science and Technology, Shanghai Institute of Optics and Fine Mechanics, Chinese Academy of Sciences, Shanghai 201800, China}
\affiliation{School of Optoelectronics, University of Chinese Academy of Sciences, Beijing 100049, China }
\affiliation{School of Physical Science and Technology, ShanghaiTech University, Shanghai 201210,  China}

\date{\today}

\begin{abstract}
We report a synergistic enhancement of betatron radiation based on the hybrid laser and plasma wakefield acceleration scheme. Quasi-phase-stable acceleration in an up-ramp plasma density first generates GeV-energy electron beams that act as a drive beam for PWFA, which then further accelerates the witness beam to GeV energies, enhancing both photon energy and flux. With this scheme, an average flux exceeding $10^{14}$ photons per steradian above 5 keV, an order of magnitude higher than the earlier reports, along with the critical energy $71 \pm 8$ keV, and a full width at half maximum divergence $(6.1 \pm 1.9)\times(5.8\pm 1.6) $ mrad$^2$ of betatron radiation were experimentally obtained. Quasi-three-dimensional particle-in-cell simulations were used to model the acceleration and radiation of the electrons in our experimental conditions, establishing a new paradigm for compact collimated hard X-ray sources.
\end{abstract}

\maketitle

Betatron radiation is generated by relativistic electrons that undergo betatron oscillations in the wakefield driven by an intense laser \cite{Rousse2004,Kneip2008}. Characterized by femtosecond pulse duration \cite{TaPhuoc2007,Horn2017}, low divergence \cite{BjrklundSvensson2021,Wang2013}, micron source size \cite{Kneip2010,Shah2006} and broadband spectra \cite{Cipiccia2011,Corde2013}, it offers unique advantages for industrial manufacturing \cite{Vargas2019,Hojbota2023}, high-energy-density physics diagnostics \cite{Mahieu2018} and biomedicine applications \cite{Albert2016,Cole2018,Wenz2015}. Recent breakthroughs in laser-plasma acceleration have led to significant enhancements in betatron radiation \cite{Kozlova2020,Yu2018,Dpp2017,Yan2014,Shou2022}. However, betatron radiation in a laser wakefield accelerator (LWFA) faces two contradictory limitations: The acceleration length is constrained by laser diffraction, pump depletion length $L_{\mathrm{pd}}^{3D} \propto n_{\mathrm{e}}^{-1}$, and dephasing length $L_{\mathrm{d}}^{3D} \propto n_{\mathrm{e}}^{-3/2}$ \cite{Lu2007}, limiting the electron Lorentz factor $\gamma$; While reducing plasma density $n_{\mathrm{e}}$ extends acceleration length,  the betatron oscillation frequency $\omega_{\beta} \propto \sqrt{n_{\mathrm{e}}/\gamma}$ diminishes, limiting the critical energy and flux of betatron radiation \cite{Corde2013}. These limitations degrade the signal-to-noise ratio and spatial resolution in imaging applications.

Recently, an innovative two-stage plasma structure combining low- and high-density regions has been proposed to significantly enhance betatron radiation characteristics. For example, the laser-driven two-stage accelerator first boost electrons to multi-GeV energies in low-density plasma, then the laser and electron beams entered into high-density plasma generating a peak brightness above $10^{26}$ photons/(s mm$^2$ mrad$^2$ 0.1$\%$BW) \cite{Zhu2020}. The hybrid laser and plasma wakefield acceleration (L-PWFA) scheme used laser-driven electron beams to drive radiator, generating MeV radiation with $>1\%$ laser-to-photon efficiency \cite{Ferri2018}. Although these simulations provide crucial theoretical foundations for radiation enhancement through two-stage plasma structures, experimental realization remains challenges in precise plasma structuring.

In this paper, we propose and experimentally demonstrate an L-PWFA scheme based on density gradient control, achieving optimization of betatron radiation. The quasi-phase-stable acceleration in an up-ramp plasma density makes it possible to produce GeV-energy electron beams. When the laser could no longer sustain LWFA, the first bubble’s electron bunch acts as the drive beam to excite PWFA. The PWFA further accelerates the witness beam, which initially exhibits large-amplitude oscillations, to GeV energies. This scheme utilizes high-energy electron beams to generate highly collimated hard X-rays, achieving an average flux exceeding $10^{14}$ photons per steradian (phs/sr) above 5 keV. Numerical simulations of the electron dynamics and radiation generation are performed to support the analysis of experimental observations.

\begin{figure}
    \centering
    \includegraphics[width=1\linewidth]{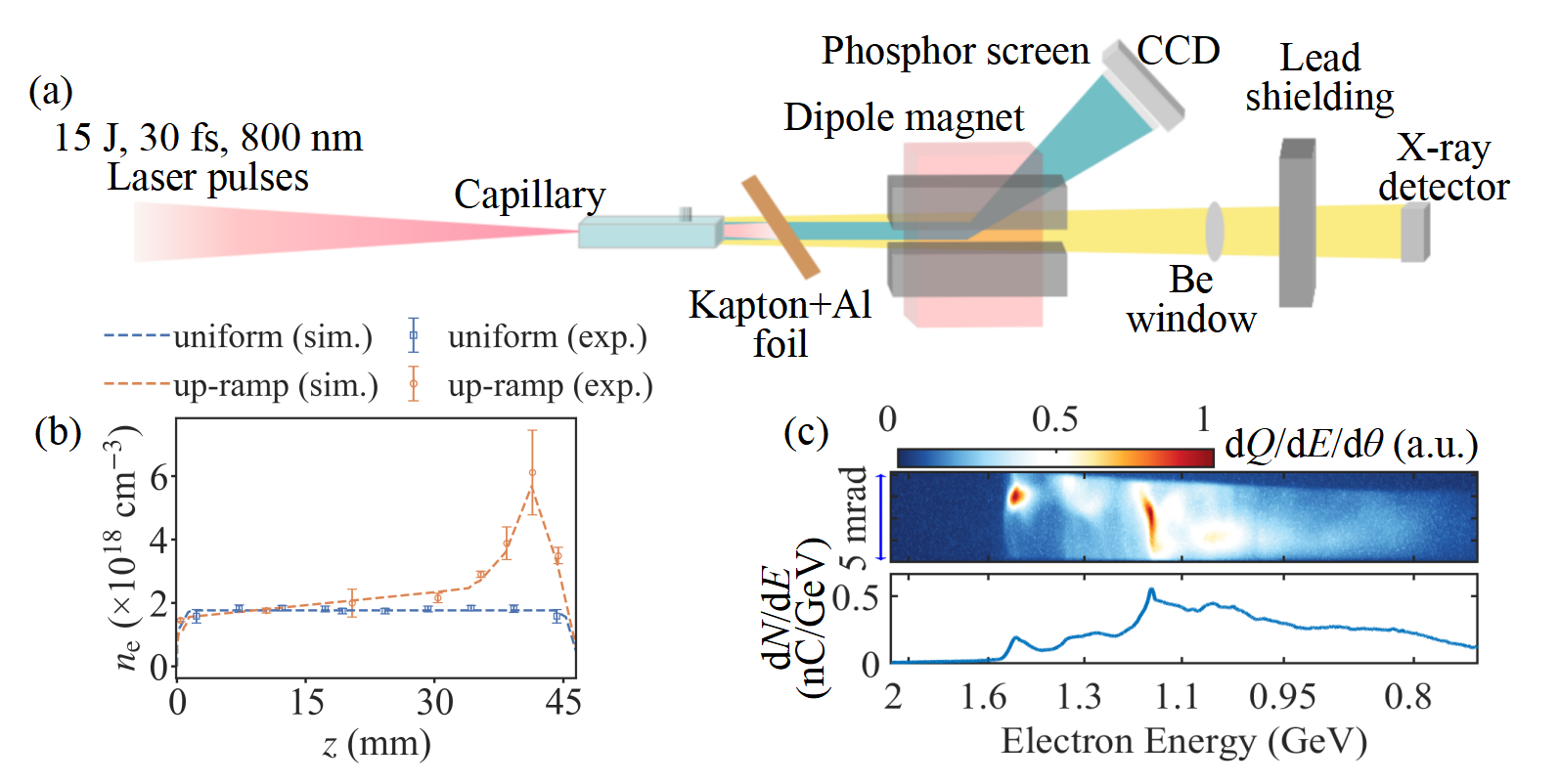}
    \caption{\label{fig:fig1}
Experimental setup. (a) An $f$/50 OAP focuses the laser pulse (red) at the entrance of the 4.5 cm long plasma capillary, producing high-energy electron (cyan) and X-ray beams (yellow). (b) Plasma density profiles in experiment and simulation (blue: uniform; orange: up-ramp). Error bars represent root mean square (RMS) of experimental measurements, dashed lines indicate simulated profiles. (c) Typical energy spectrum of the electron beam generated by L-PWFA in the up-ramp plasma density. 
 }
\end{figure}

The experiment was conducted using the 1 PW/0.1 Hz laser system at the Shanghai Superintense Ultrafast Laser Facility (SULF) \cite{Zhang2020}. The experimental setup is shown in Fig.~\ref{fig:fig1}(a). The linearly polarized laser pulse (800 nm, 15 J, 30 fs) was focused using an $f$/50 off-axis parabolic mirror onto the capillary entrance. The $1/{\mathrm{e}}^2$ radius of focal spot was 51 μm, with 66\% of the energy contained within its intensity contour. The peak intensity reached $7.0 \times 10^{18} ~\mathrm{W}/\mathrm{cm}^2$, corresponding to a normalized vector potential of $a_0=1.8$.

The capillary was filled with pure He gas and the plasma density was measured by Stark broadening analysis \cite{Qin2018} of the He I 587.6 nm emission line, with results presented in Fig.~\ref{fig:fig1}(b). Radiation characteristics were optimized for uniform plasma density (maintained at $3.6\times10^{18} $ cm$^{-3}$) and up-ramp plasma density (varying from 1.2 to $5.6\times10^{18}$  cm$^{-3}$) configurations. See Appendix A for details on capillary design and plasma density measurements.

The depleted laser was blocked by a composite foil (50-μm-thick Kapton + 50-μm-thick Al) after the interaction, which simultaneously served to suppress X-ray photons below 5 keV. The X-ray beam exited the vacuum chamber through a 300-μm-thick Be window located 4.6 m downstream from the source, then passed through an aperture in the lead shielding wall which effectively suppresses bremsstrahlung interference from deflected electron beams. The X-ray beam was ultimately detected by a detector consisting of a 400-μm-thick Gd$_2$O$_2$S:Tb scintillator directly coupled to a CMOS chip, with an effective pixel size of 49.5 μm. The electron beam was characterized using a magnetic spectrometer system located 4 meters downstream from the capillary exit, which included a 180-cm-long electromagnetic dipole with peak magnetic fields up to 1.5 Tesla, a Lanex scintillating screen, and a 14-bit CCD camera for beam imaging, with 5 mrad vertical angular acceptance. Figure~\ref{fig:fig1}(c) shows a typical energy spectrum of the electron beam with a maximum energy of 1.5 GeV, containing $> 200$ pC of charge counted in the 0.8-2.5 GeV energy range. 
\begin{figure}
    \centering
    \includegraphics[width=1\linewidth]{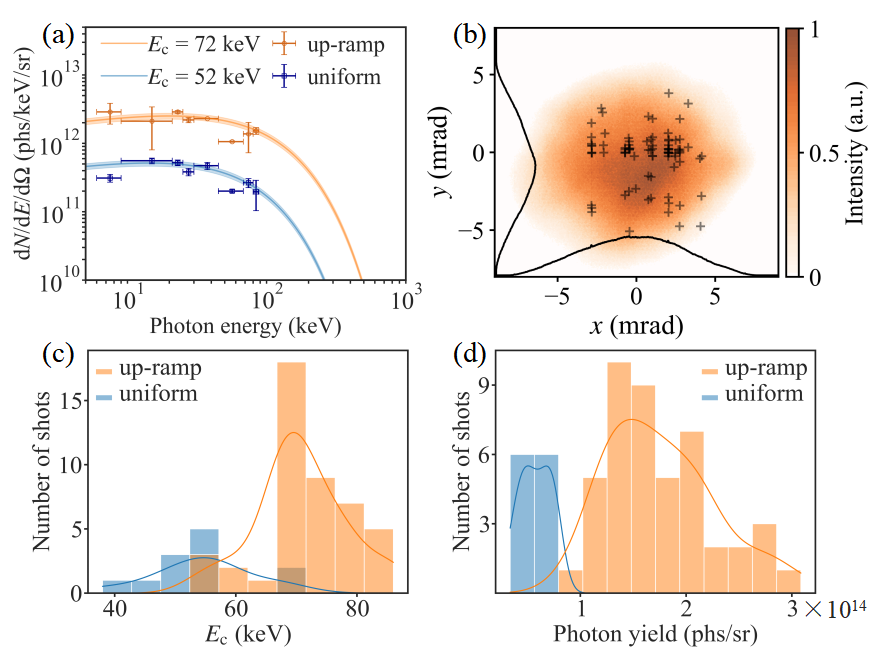}
    \caption{\label{fig:fig2}
Spectral and statistical properties of betatron radiation. (a) Measured betatron spectra comparing uniform (blue) and up-ramp (orange) plasma density, with error bars showing the RMS variation across three shots. Synchrotron model fits (solid lines) include $\pm 10\%$ intensity uncertainty bands (shading). See Appendix D for details of Ross filter pairs. (b) Single-shot angular distribution of the radiation in the up-ramp plasma density and statistical distribution of the radiation central positions over 85 consecutive shots. (c, d) Statistical distributions of critical energy and photons per steradian for uniform (12 shots) and up-ramp (45 shots) plasma density.
 }
\end{figure}

The radiation characteristics of the two structures were studied using Ross filter pairs, and the results were fitted with synchrotron spectra as shown in Fig.~\ref{fig:fig2}(a). Under the up-ramp plasma density structure, the signal intensities of the filters were enhanced $3-4$ times compared to the uniform plasma density structure, with the critical energy increasing from 52 keV to 72 keV. The statistical results are shown in Figs.~\ref{fig:fig2}(c) and~\ref{fig:fig2}(d). For the uniform plasma density case, the average critical energy of betatron radiation was 55 keV (maximum 67 keV), with an average photon yield of $5.8\times10^{13}$ phs/sr (maximum $7.6\times10^{13}$ phs/sr). For the up-ramp plasma density case, the average critical energy increased to 71 keV (maximum 86 keV), and the average photon yield rose to $1.8\times10^{14}$ phs/sr (maximum $3.1\times10^{14}$ ps/sr) above 5 keV. Overall, the up-ramp plasma density structure enhanced the critical energy by a factor of 1.8 and the photon yield by a factor of $3-4$. As typical radiation divergence of the up-ramp plasma density configuration shown in Fig.~\ref{fig:fig2}(b), measurements revealed a full width at half maximum (FWHM) divergence of $(6.1\pm1.9)\times(5.8\pm1.6)$ mrad$^{2}$, where the values represent mean ± standard deviation ($\sigma$) in the horizontal and vertical directions. Statistical analysis reveals horizontal and vertical centroid fluctuations with $\sigma_x=2.0$ mrad and $\sigma_y=1.8$ mrad, respectively. Considering the $\pm 3\sigma$ range of the FWHM divergence and combining the spectral distribution of the up-ramp plasma density structure, the average photons per shot exceeded $3.3\times10^{10}$ ($> 5$ keV), with a maximum of $5.6\times10^{10}$ ($> 5$ keV).

FBPIC simulations \cite{Lehe2016,Jalas2017} were performed to investigate how the L-PWFA mechanism in the up-ramp density can enhance the critical energy and yield of betatron radiation. The simulations  were set to match the experimental parameters. The simulation window had longitudinal and transverse dimensions of 120 μm and 135 μm, respectively. Each grid cell contained 16 macro-particles, with cell sizes of $\Delta z = 60$ nm and $\Delta r = 135$ nm. A polarized laser with a wavelength of $\lambda_0 = 0.8$ μm, a normalized intensity of $a_0 = 1.8$, a pulse duration of $\tau = 30$ fs, and a focal spot size of $\omega_0= 51$ μm was launched from the left side of the simulation window, propagating along the $z$-direction and focusing at a position 1.25 mm from the plasma entrance. The longitudinal distribution of the plasma density is shown by the orange dashed line in Fig.~\ref{fig:fig1}(b). 

\begin{figure*}
    \centering
    \includegraphics[width=0.9\linewidth]{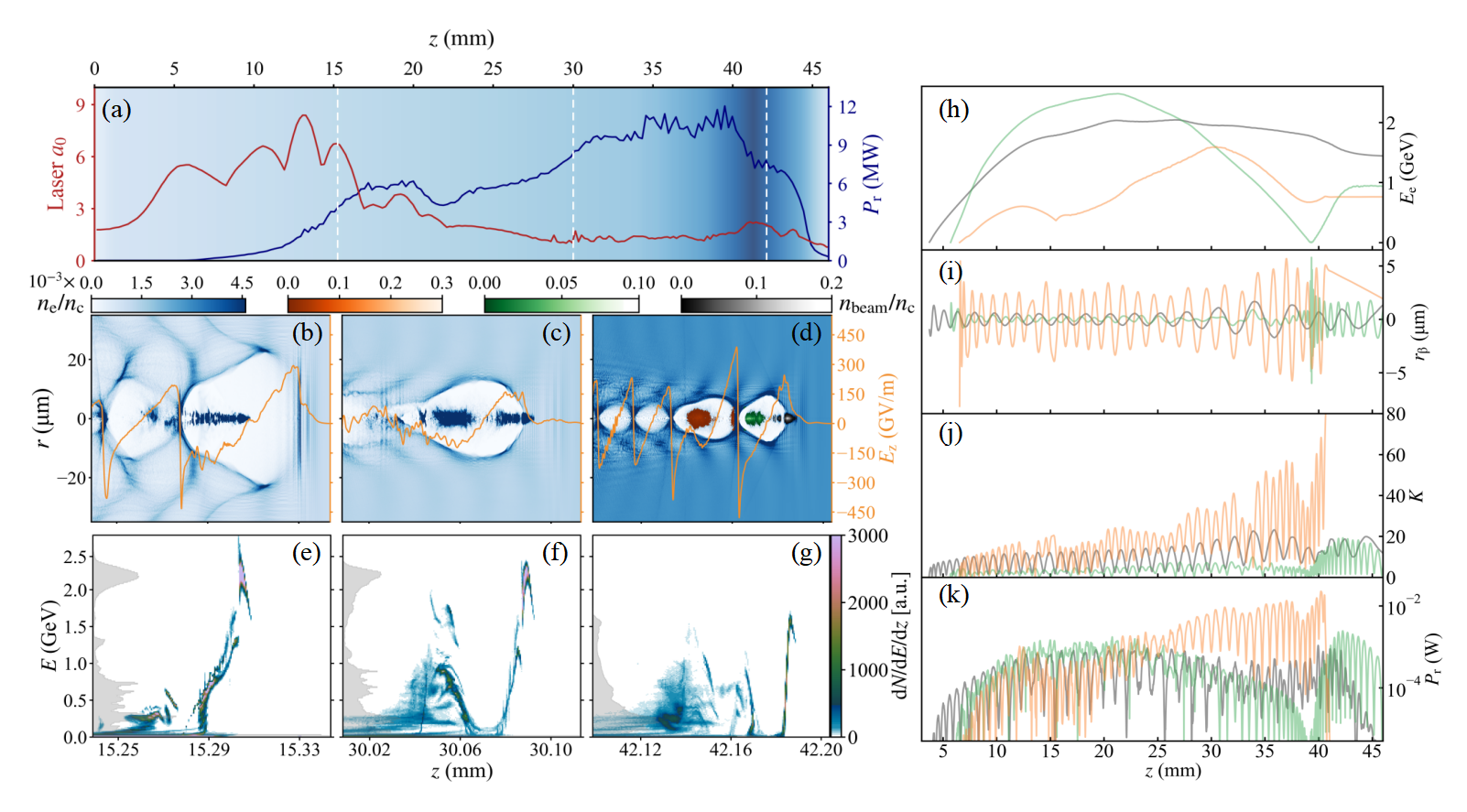}
    \caption{\label{fig:fig3}
    Electron beam dynamics from cascaded L-PWFA in PIC simulations. (a) Spatiotemporal evolution of plasma density, the normalized intensity $a_0$, and instantaneous betatron radiation power $P_{\mathrm{r}}$, with three white dashed lines indicating times corresponding to (b)$-$(d). (b)$-$(d) Plasma density distributions in $r-z$ plane with corresponding on-axis plasma field (orange curves), and $n_{\mathrm{c}}$ represents a critical plasma density. (b) Pure LWFA stage, showing electron beam acceleration in the first and second bubbles; (c) Transition from LWFA to PWFA, where the first-bubble electron beam becomes the drive beam while the second-bubble beam serves as witness beam; (d) At increased density, the witness beam spans two PWFA bubbles and splits into two bunches. (e)$-$(g) Corresponding longitudinal phase spaces of electron beams and gray shading represents energy spectra. (h)$-$(k) Spatiotemporal evolution of single-electron parameters: energy, oscillation amplitude $r_{\beta}$, strength parameter $K$, and $P_{\mathrm{r}}$. The orange, green, and gray curves correspond to the representative electrons from the three electron bunches marked in (d).
    }
\end{figure*}

Figure~\ref{fig:fig3}(a) reveals the entire process of enhanced betatron radiation under the L-PWFA mechanism. During the first 17 mm of the simulation, LWFA dominated. The linearly increasing plasma density in this region was $1.2-1.9\times10^{18}$ cm$^{-3}$, higher than the typical LWFA range of $3-8\times10^{17}$ cm$^{-3}$ \cite{Leemans2014,Gonsalves2019,Picksley2024}.As the laser self-focused in the plasma and the bubble expanded, electrons were continuously injected into the bubble \cite{Kuschel2018}. The electrons were further accelerated in the accelerating phase, continuously gaining energy, which progressively increased the total betatron radiation power. As shown in Figs.~\ref{fig:fig3}(b) and~\ref{fig:fig3}(e), within the first 15.3 mm of the plasma, the electron beam in the first bubble was maintained in the accelerating phase, with its energy reaching above 2.5 GeV. In contrast, electrons in the secondary bubble experienced significant slippage into the decelerating phase due to the expansion of the first bubble, consequently limiting their maximum energy to $\sim 0.5$ GeV.

Within the $17-22.5$ mm range inside the plasma, laser defocusing caused the decay of $a_0$, triggering a transition from LWFA to a hybrid L-PWFA regime. As the pump energy depleted, the laser could no longer sustain stable bubble structures, resulting in multi-bubble merging. During this process, some electrons from the tail of the electron beam slipped from the accelerating phase of the first bubble to the decelerating phase of the second bubble, resulting in a rapid energy loss. While betatron radiation power remained initially high due to persistent LWFA effects, its intensity diminished progressively as the LWFA contribution weakened.

In LWFA, the bubble size $r_{\mathrm{b}}\propto \sqrt{a_0/n_{\mathrm{e}}}$ decreases with declining laser intensity and increasing plasma density \cite{Corde2013}. Beyond 22.5 mm in the plasma, the bubble expanded to dimensions significantly exceeding those driven by the laser, marking the PWFA-dominated regime \cite{Gtzfried2020}. The electron beam density in the first bubble satisfied the condition $n_{\mathrm{beam}} > 1.8n_{\mathrm{e}}$ \cite{Lu2006}, making it sufficient to act as the drive beam for PWFA. The electron beam in the second bubble served as the witness beam in the PWFA. The betatron radiation power gradually increased with the energy gain of the witness beam. As shown in Figs.~\ref{fig:fig3}(c) and~\ref{fig:fig3}(f), at 30 mm the drive beam's maximum energy decreased below 2.5 GeV, while the witness beam increased beyond 1.5 GeV. Beyond this point, the increasing plasma density induced bubble contraction, causing the witness beam to span two PWFA bubbles: electrons in the tail region of the first bubble gained energy, while those in the head region of the second bubble experienced energy loss. As depicted in Fig.~\ref{fig:fig3}(d) and~\ref{fig:fig3}(g), the density of the drive beam, the witness beam in the first bubble, and the witness beam in the second bubble are represented by the grayscale, green, and orange colorbars, respectively. The maximum energy of the drive beam decreased to 1.6 GeV, while the witness beams in the first and second bubbles were evolved to 1.3 GeV and 1.5 GeV, respectively. The simulation results show that the electron beam energy spectrum and charge ($\sim 377$ pC) in the range of $0.8-2$ GeV are in good agreement with the experimental results presented in Fig.~\ref{fig:fig1}(c).

To better investigate each electron bunch's contribution to betatron radiation, three representative electrons [color-coded in gray, green, and orange in Fig.~\ref{fig:fig3}(d), corresponding to the drive and two witness bunches respectively] were analyzed by tracking their key parameter evolutions: energy, oscillation amplitude $r_{\beta}$, strength parameter $K=r_{\beta}\omega_{\beta}\gamma$/c, and instantaneous radiation power $P_{\mathrm{r}}$, as shown in Figs.~\ref{fig:fig3}(h) $-$~\ref{fig:fig3}(k). The electrons in the gray beam were captured and stably maintained at the tail of the bubble in the up-ramp density gradient region of the plasma during the LWFA, continuously gaining energy through acceleration. Due to the relatively low plasma density in this phase, $r_\beta$ was confined to approximately 1 μm. As these high-energy electrons acted as the drive beam to excite the PWFA stage, their energy was progressively transferred to the plasma wakefields. With an increase in plasma density, the energy loss rate of the drive beam accelerated significantly, and $r_\beta$ increased from 1 μm to 3 μm. Remarkably, despite variations in both $\gamma$ and $r_\beta$, $K$ and $P_{\mathrm{r}}$ remained relatively stable due to their compensatory changes.

The electrons in the first witness beam (green) exhibited delayed self-injection during the LWFA stage, where progressive laser focusing and bubble expansion enabled their eventual trapping and acceleration to high energies. Upon transition to the PWFA stage, these electrons, initially positioned at the rear of the drive beam within the bubble's decelerating phase, first underwent energy depletion to zero. However, subsequent plasma density increases induced rapid bubble contraction, causing the electrons to slip back into an accelerating phase and regain substantial energy. The high-density plasma environment significantly increased the betatron oscillation frequency $\omega_{\beta}$ of the electrons, leading to a notable enhancement in $K$ and $P_{\mathrm{r}}$.

The electrons in the second witness beam (orange), originating from the secondary bubble of the LWFA stage, developed large-amplitude oscillations via transverse injection \cite{Corde-2013}. However, inter-bubble interactions caused slippage into a decelerating phase, ultimately limiting energy gain. When entering the PWFA stage, these electrons partially recovered energy through the high-density drive beam's wakefield, though the strong beam-loading effect diminished the effective acceleration gradient. Initially, $K$ and $P_{\mathrm{r}}$ increased with the electron energy gain. As the bubble contracted due to the rapid growth of plasma density, these electrons slipped into a decelerating phase, losing energy while $K$ and $P_{\mathrm{r}}$ continued growing through enhanced betatron oscillation amplitude $r_\beta$ and frequency $\omega_\beta$. With a remarkable charge exceeding 1 nC, nearly an order of magnitude higher than both the drive beam and the first witness beam, they dominated the radiation output, elevating the total power to 10 MW in the PWFA regime.

\begin{figure}
    \centering
    \includegraphics[width=1\linewidth]{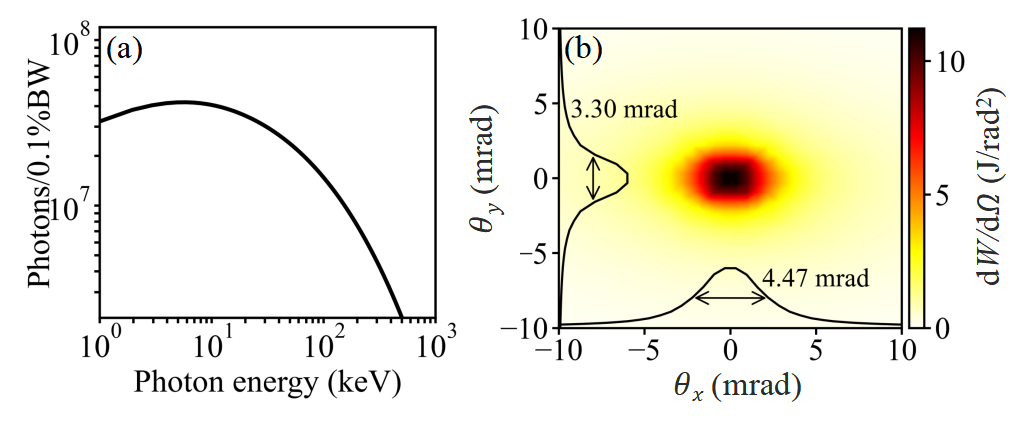}
    \caption{\label{fig:fig4}
   (a) Betatron radiation spectrum. (b) Angular distribution of betatron radiation, with the line indicating the axial distribution and the label representing FWHM divergence angle.
 }
\end{figure}

The characteristics of betatron radiation in the simulation were investigated by post-processing the electron trajectories using the SynchRad code \cite{Andriyash} based on the Fourier transform of the Lienard-Wiechert potentials \cite{Esarey2002}. See Appendix D for more details. Figure~\ref{fig:fig4}(a) presents the simulated X-ray spectral distribution of betatron radiation generated via the L-PWFA mechanism in the up-ramp density profile, featuring an on-axis peak at 10 keV extending up to 500 keV. The spectrum exhibits a critical energy $E_{\mathrm{c}} = 90$ keV and a total photon yield of $9.6\times10^{10}$ ($>5$ keV), showing good agreement with experimentally measured photon yields. The total energy of the emitted photons can reach 735 μJ. Figure~\ref{fig:fig4}(b) shows the angular distribution of radiation, with an FWHM divergence angle of $4.17\times3.30$ mrad$^2$. This was close to the minimum divergence angle measured in the experiment. According to the simulations, with an electron beam FWHM size of 40 μm in length and 3 μm in width, the peak brightness in the experiment can be estimated as $8\times10^{23}$ photons/(s mm$^2$ mrad$^2$ 0.1$\%$BW) at 30 keV. 

In conclusion, we have demonstrated the synergistic enhancement of betatron radiation based on the hybrid L-PWFA mechanism. Experimental results showed this scheme produced an average flux exceeding $10^{14}$ photons/sr above 5 keV, significantly surpassing the performance of single-stage LWFA \cite{Cole2018,Bloom2020,Cole2015,Hussein2019,Zhang2025}.The critical energy of this source was already comparable to that of high-energy synchrotron radiation facilities. Simulations revealed that this enhancement stemmed from cooperative dual-stage wakefield dynamics: using the electron beam from the first bubble of LWFA as the drive beam to excite the wakefield (PWFA) not only amplified transverse oscillations through plasma density modulation but also accelerates the witness beam to GeV energies with large oscillation amplitudes, simultaneously optimizing both the critical energy and flux of betatron radiation. This compact X-ray source enables phase-contrast imaging of laser-driven shock waves and X-ray absorption near-edge structure (XANES) spectroscopy, providing a critical tool for investigating dynamic compression, phase transitions, and non-equilibrium materials \cite{Mahieu2018,Wood2018,Gunot2022}. 

\begin{acknowledgments}
Special thanks to the staff at SULF who provided good conditions and operation. This work was supported by the National Natural Science Foundation of China (Grant Nos. 12388102, 12225411, 12474349 and 12174410), the Strategic Priority Research Program of the Chinese Academy of Sciences (Grant No. XDB0890201 and XDB0890202), CAS Project for Young Scientists in Basic Research (Grant No. YSBR060), and CAS Youth Innovation Promotion Association (No. 2022242). 

\end{acknowledgments}

\bibliography{ref}

\appendix
\section{\label{app:A}Appendix A: Measurement of plasma density}
The gas target configuration incorporated two capillary sections to generate tailored plasma density profiles: a 30-mm-long tapered capillary with inner diameter linearly decreasing from 3 mm to 2 mm to produce an increasing plasma density gradient, and a 10-mm-long uniform capillary with constant diameter 1.5 mm to create a high-density. Pure He gas was injected through a 5-mm inter-capillary gap at 3.4 bar gas pressure. In the comparative experiments, a uniform plasma density distribution was achieved by using a capillary with a pore size of 2.5 mm. The plasma density within the capillary was determined through spectral line broadening analysis of the He I 587.6 nm emission line, with the measured data points presented in Fig.~\ref{fig:fig1}(b). 

\section{\label{app:B}Appendix B: Comparison of electron beams at two plasma density distributions}
\begin{figure}
    \centering
    \includegraphics[width=0.8\linewidth]{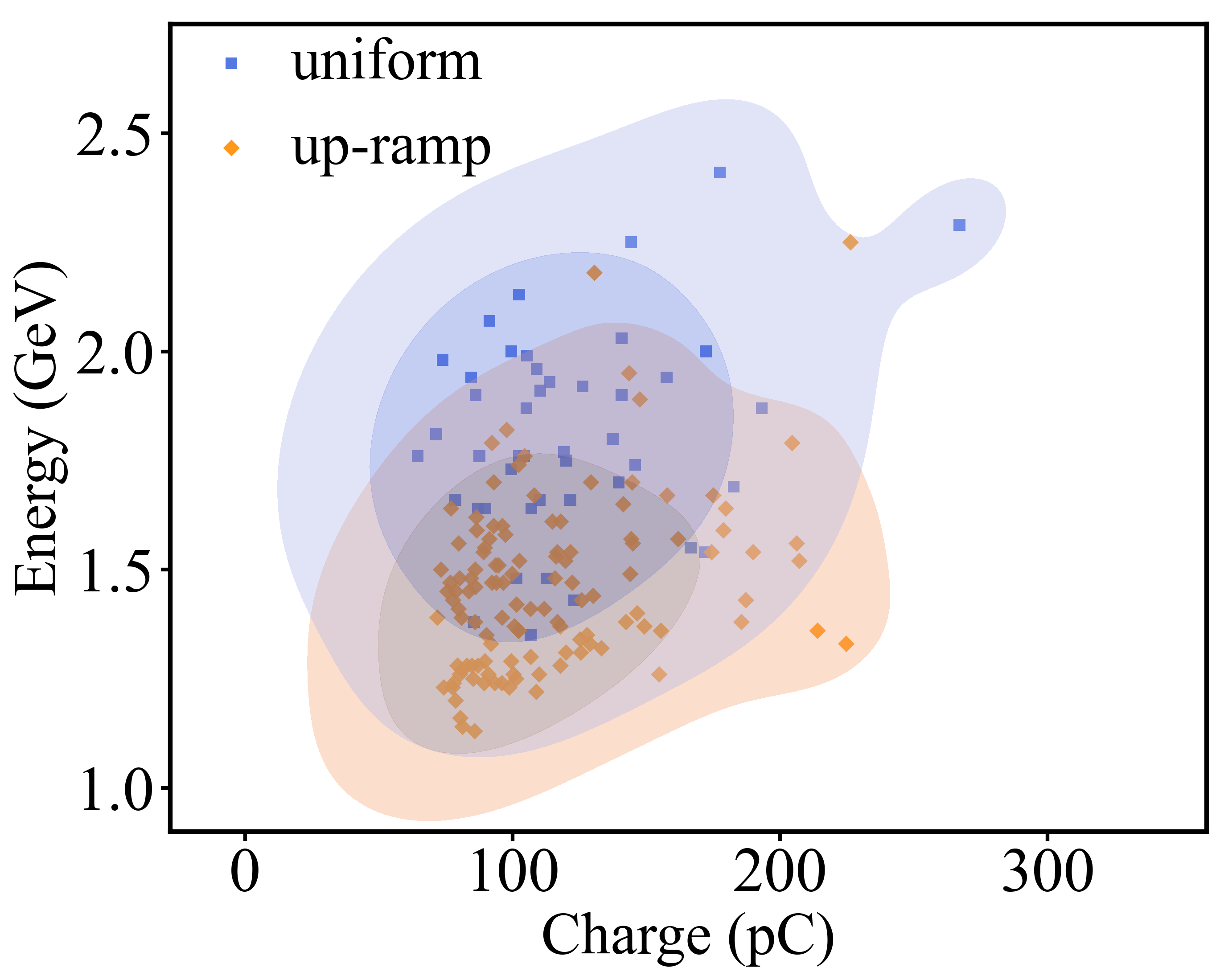}
    \caption{\label{fig:fig5}
    Statistical comparison of electron beam energy and charge under the uniform (45 shots) and the up-ramp (120 shots) plasma density distributions. 
 }
\end{figure}
The characteristics of electron beams generated under two plasma density distributions are shown in Fig.~\ref{fig:fig5}. The darker shaded regions represented the $1\sigma $ $(68.3\%)$ confidence intervals of the core distribution, while the lighter shading corresponded to extended $2\sigma$ $(95.4\%)$ coverage. The charge of the electron beam generated by the two structures was concentrated in the range of $70-200$ pC, and the maximum energy of the electron beam generated by the up-ramp plasma density was significantly lower than that of the uniform plasma density. This was because during the PWFA process in the up-ramp plasma density, the energy of the driving beam was transferred to the plasma wakefields, resulting in a significant decrease in energy, and the electron transverse oscillation amplitude and oscillation frequency of the witness beam were large. Although it limited the energy gain of the electron beams, it was beneficial in enhancing betatron radiation. In contrast, the uniform plasma density enabled sustained electron acceleration via LWFA, with slower energy loss after laser depletion, so higher energy can be obtained eventually, but it did not significantly enhance betatron radiation.

\section{\label{app:C}Appendix C: Material and thickness of filters}
\begin{figure}
    \centering
    \includegraphics[width=1\linewidth]{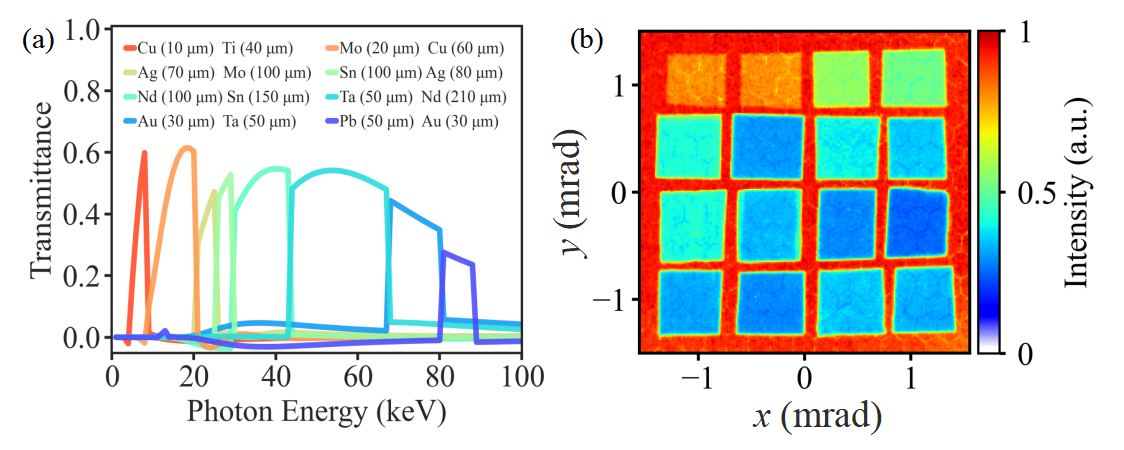}
    \caption{\label{fig:fig6}
    (a) Transmittance of each filter pair in the Ross filter. (b) The raw data of X-rays attenuated by Ross filters.
 }
\end{figure}

 The energy spectra of both configurations were reconstructed from three-shot experimental datasets using the Ross filter pair method. The materials and thicknesses of the filter pairs are shown in Fig.~\ref{fig:fig6}, and their transmissions were calculated using the XOP software toolkit \cite{RN271}. Therefore, the X-ray energy spectrum can be expressed as:
 \begin{eqnarray}
\frac{{\mathrm{d}}N}{{\mathrm{d}}E{\mathrm{d}}\Omega}=\frac{\Delta S}{C\cdot\Delta E\cdot {\mathrm{d}}\Omega},
\end{eqnarray}
\label{eq:1}where $\Delta S$ is the intensity difference of the filter pair measured by the detector, $C$ is the system's response to the filter pair, which includes the detector's response to photons and the attenuation effects of 10 cm of air, a 50 μm Kapton foil, a 50 μm Al foil, and a 300 μm Be window. $\Delta E$ represents the energy bandwidth of each filter pair, and $d\Omega$ is the solid angle subtended by the filter pair. The measured results were fitted with a synchrotron radiation spectrum to obtain the comparative radiation characteristics under the two density structures, as shown in Fig.~\ref{fig:fig2}(a).

\section{\label{app:D}Appendix D: Computation of the betatron source.}
The angular-spectra distributions of betatron radiation was computed from 20,000 representative electron trajectories with a minimal energy of 5 MeV using the following formula: 
\begin{eqnarray}
\frac{{\mathrm{d}}^2W}{{\mathrm{d}}\omega {\mathrm{d}}\Omega} = \frac{e^2}{16\pi^3 \epsilon_0 c} \left| \int_{-\infty}^{\infty} e^{i\omega \left( t - \frac{\mathbf{n} \cdot \mathbf{r}(t)}{c} \right)} \cdot\mathcal{F}(t)  \, {\mathrm{d}}t\right|^2,
\end{eqnarray}
\begin{eqnarray}
\mathcal{F}(t) =\frac{
    \mathbf{n} \times \left[ (\mathbf{n} - \boldsymbol{\beta}(t)) \times \dot{\boldsymbol{\beta}}(t) \right]
}{\left( 1 - \boldsymbol{\beta}(t) \cdot \mathbf{n} \right)^2},
\end{eqnarray}
where \textbf{r} is the radius vector of the electron, and \textbf{n} is the unit vector of observation. This formula was integrated over a $40 \times 40$ mrad$^2$ solid angle to obtain the spectrum presented in Fig.~\ref{fig:fig4}(a). The instantaneous radiated power can be calculated by the relativistic generalization of Larmor’s formula:
\begin{eqnarray}
P_{\mathrm{r}} = \frac{2 e^{2}}{3 c} \, \gamma^{6} \, \left( \dot{\boldsymbol{\beta}}^{\, 2} - \left[ \boldsymbol{\beta} \times \dot{\boldsymbol{\beta}} \right]^{\, 2} \right), 
\end{eqnarray}
where $\boldsymbol{\beta}=\textbf{v}/c$ is the dimensionless velocity vector, and $\gamma$ is the Lorentz factor of the electron.

\end{document}